
%
%
\magnification=\magstep1
\baselineskip=12pt

\def\u{{\bf u}}
\font\Mid=cmr10 scaled \magstep2

\def\\{\hfill\break}
\centerline {\bf\Mid Transition to chaos in a shell model of turbulence.}
\hfill\break
\hfill\break
\hfill\break
\centerline{L. Biferale$^{1}$,
 A. Lambert$^{2}$, R. Lima$^{2,3}$ and G. Paladin$^{4}$}
\bigskip
\noindent
\item{$^1$}
{\it Observatoire de Nice,  B.P. 229, 06304 Nice Cedex 4,  France}
\noindent
\item{$^2$}
{\it Centre de Physique Th\'eorique, CNRS - Luminy
\hfill\break
 case 907, F-13288  Marseille, Cedex 9,  France}
\noindent
\item{$^3$}
{\it The Benjamin Levich Institute,
 The City College of the City University of New York
\hfill\break
Steinman Hall, 140th St. and Convent av.
New York, NY 10031, USA}
\item{$^4$}{\it Dipartimento di Fisica, Universit\`a dell'Aquila,\hfill\break
Via Vetoio, Coppito  I-67100 L'Aquila, Italy}
\bigskip
\bigskip
\bigskip
\centerline {ABSTRACT}
\medskip
We study a shell model
 for the energy cascade in three dimensional turbulence
 at varying the coefficients of the non-linear terms in such a way that
 the fundamental symmetries of Navier-Stokes are conserved.
When a control parameter $\epsilon$ related to the  strength of
 backward energy transfer is enough small, the dynamical system
 has a stable fixed point corresponding to the Kolmogorov
 scaling.  This point becomes unstable at $\epsilon=0.3843...$
 where a stable limit cycle appears  via a Hopf bifurcation.
 By using the bi-orthogonal decomposition,
 the transition to chaos is shown to follow the Ruelle-Takens scenario.
For $\epsilon > 0.3953..$ the dynamical evolution is intermittent
 with a positive Lyapunov exponent.
 In this regime, there exists a
 strange attractor which remains close to the Kolmogorov (now unstable)
  fixed point, and a local scaling  invariance which
  can be described via a  intermittent one-dimensional map.
\vfill\eject
\noindent
{\bf { Introduction} }
\bigskip
Shell models for the energy cascade in fully developed turbulence
were introduced to mimic the Navier-Stokes equations
$$
\partial_t \u \  + \ ({\bf u} \cdot \nabla) \u \ = \
 - { {\bf \nabla} P \over \rho } +
 \nu \Delta \u \ + {\bf f}_u .
\eqno(1.1)
$$
The reason is that, in a turbulent regime,  the
number of  degrees of freedom necessary to describe the flow generated
 by eq.s (1.1) is enormous since it
 roughly increases  as  a power of the Reynolds number,
 $Re^{9/4}$.  However, these degrees of freedom probably are
 organized in a hierarchical way, so that one expects that
 simplified dynamical systems  could be relevant for the description
of the scaling invariance.
The basic idea of shell models
is to consider a {\it discrete} set of wavevectors ,
`shells', in $k$-space, and to construct
an ordinary differential equation on each shell. The
 form of the coupling terms among the
various shells are chosen
 according the main symmetries of the Navier Stokes equations.

Standard shell models have a relatively small number of degrees of
freedom, so that they can be analyzed  as a  dynamical system [1-5].

The set of ODE are derived under the assumption
 that the most relevant mechanism for the behaviour of the velocity field,
$u$, is given by a cascade transfer from large to small scales.

Among the huge literature existent nowadays on shell models
one finds interesting results about the properties of static solutions
for the  Novikov-Desnynasky model [3,5], or numerical and
analytical studies of the GOY
(Gladzer, Ohkitani and Yamada) model in the strongly chaotic regime [1,2,6,8].
  Amazingly enough, there are
not detailed studies of the transition to chaos presents in all of these
models.
 The aim of this paper is to fill this gap, by studying this issue in the
case of the GOY model. The understanding of the ``route to chaos''
 is of primary importance to enlighten the intermittent
character of the dynamics. In fact, we shall see that in the chaotic
regime immediately above the transition, we can study the GOY model
 through a one-dimensional map that captures the main
 dynamical mechanisms which are of the origin  of intermittency.
 These mechanisms are much more difficult to be analyzed
 in the fully chaotic regime which is usually considered.
  However, it is an open problem
to understand  how much the dynamical intermittency of
 shell models is a realistic approximation of
 real turbulence.

The paper is organized as follows:

In section 1 we define the GOY model. In
section 2 there is a detailed discussion on the relative importance of
the forward to backward transfer of energy. In section 3 we present our
numerical results on the transition to chaos, analyzed by using a
bi-orthogonal decomposition [7]; in section 4 we introduce an ad-hoc modified
GOY model which allows us to perform a detailed analysis of the intermittent
properties nearby the transition.

All the numerical integrations presented hereafter are performed by
using a (second-order) slaved Adam-Beshforth scheme [6].
 Results  for the model with $19$ shells have been obtained by choosing
 a time step of $\delta t= 3 \, 10^{-4}$ ,
 $\nu=10^{-6}$,
$f=5 \, 10^{-3} \times (1+i)$ and $k_0=0.05$, while for the model with
$27$ shells: $\delta t= 10^{-5}$ ,
 $\nu=10^{-9}$,
$f=5 \, 10^{-3} \times (1+i)$ and $k_0=0.0625$.
\bigskip
\medskip
\noindent
{\bf 1. The GOY model}
\bigskip

 In shell models the Fourier space is divided in $N$ shells,
 each shell $k_n$ ($n=1,2,..,N$)
 consisting of the wavenumbers with modulus
  $k$ such that $k_0 2^n < k < k_0 2^{n+1}$.  The velocity
 increments  $|u(\ell)-u(x+\ell)|$ on scale $\ell \sim k_n^{-1}$
 are given by the complex variables
 $u_n$. The evolution equations are obtained according the following criteria:
\item{(a)} the linear term for $u_n$ is given by $-\nu k_n^2 u_n$
\item{(b)} the non-linear terms for $u_n$ are combination of the form
        $k_n u_{n'} u_{n''}$
\item{(c)} the interactions among shells are local in $k$-space
 (i.e. $n'$ and $n''$ are close to $n$)
\item{(d)} in absence of forcing and damping
one has conservation of volume in phase space
and the conservation of energy ${1\over 2} \,
 \sum_n |u_n|^2$.

In the GOY model the shells
$n'$ and $n''$ are nearest and next nearest neighbors of $n$ so that
the evolution  equations are:
$$
({d\over dt}+\nu k_n^2 ) \ u_n \ =
 i \, k_n \, (a_n \, u^*_{n+1} u^*_{n+2} \, + \, {b_n\over 2}  \,
u^*_{n-1} u^*_{n+1} \, + \,
 {c_n \over 4} \,  u^*_{n-1} u^*_{n-2})  \ + \ f \delta_{n,4},
\eqno(1.2)
$$
 with $n=1,\cdots N$ and  boundary conditions
$$
b_1=b_N=c_1=c_2=a_{N-1}=a_N=0.
\eqno(1.3)
$$
The velocity $u_n$ is a complex variable, $\nu$ is the viscosity,
 and
 $f$  is an external  forcing  (here on  the fourth mode).
The  coefficients of the non-linear terms  should obey the relation
 $$
a_n+b_{n+1}+c_{n+2}=0
\eqno(1.4)
$$
 to satisfy the conservation  of
 $\sum_n |u_n|^2$ (energy) in the  absence of  forcing
 and with $\nu = 0$.
 Moreover,  they are defined modulus a multiplicative factor
(related to a  time  rescaling), so that one can fix $a_n=1$.
As a consequence,
 the respect of the main symmetries of the Navier Stokes equations
 still leaves a free parameter $\epsilon$ so that
$$
a_n=1 \qquad b_n=-\epsilon  \qquad c_n=-(1-\epsilon).
\eqno(1.5)
$$
As we will see in the following, the parameter $\epsilon$ plays
an important role in defining both  static and dynamical properties
of the model.

It is important stressing that scaling law of Kolmogorov
 ($u_n\sim k_n^{-1/3}$ )
 is a   fixed point of the  inviscid  unforced
evolution equations with $N \to \infty$ and neglecting
 the infrared boundary conditions (1.3). In the next sections
 we show that the Kolmogorov scaling remains  a fixed point of the
 shell model with $N$ shells, forcing  and finite viscosity,
 and plays a key role in the dynamics.
\bigskip
\medskip
\noindent
{\bf 2. Static and Dynamical properties}
\bigskip

The GOY model has been defined such as to have ``Kolmogorov 1941''
(K41) static-solutions in the inviscid ($ \nu=0$) unforced
limit and for the number of shells $ N \rightarrow \infty$:
  $ u_n \sim k_n^{-1/3} $.
  Actually, by studying
the static properties of the model, it is easy to recognize that there are
two infinite  sets of static solutions.
Solutions belonging to the same set are characterized by possessing
 the same scaling exponent.

For the GOY model we have  the following two possible static behaviours:

\item{(1)} Kolmogorov-like: $ u^{K41}_n = k_n^{-1/3} g_1(n) $;  with $g_1(n)$
being any
periodic function of period three.

\item{(2)} Fluxless-like: $ u^{fl}_n =  k_n^{ ( \ln_2 {|\epsilon-1| \over
2})/3}
g_2(n)$;  where still $g_2(n)$ is any periodic function of period three.

As long as one is interested in scaling laws, the presence
of superimposed periodic oscillations could seem particularly disappointing.
Nevertheless, the existence in the phase space of an infinite manifold
 K41-like, instead of a single point,
   will  turn out
to be relevant for  the dynamical properties of
the model.

In order to focus only on the power law scaling it is useful to study
the static behaviour of the ratios:
$$
q_n = u_{n+3}/u_{n}.
\eqno(2.1)
$$
Let us notice
that the same
set of observables  have already been used to describe some
exotic (chaotic) behaviours of the  energy  cascade in a different class
of shell models [10].

In terms of the $q_n$'s, a static and inviscid solution of eqs. (1.2)
  can be generated by the iterations of the following
one-dimensional complex ratio-map:

$$q_n = {\epsilon \over 2} + { (1-\epsilon) \over 4 q_{n-1}}.
\eqno(2.2)
$$

The map (2.2) has two fixed points $q^{K41},q^{fl}$
corresponding to the two possible scaling behaviours for the $u_n$'s:

\item{(1)} $q^{K41} = 1/2 \rightarrow u_n \sim u^{K41}_n$
\item{(2)} $q^{fl} = { (\epsilon -1) \over 2}  \rightarrow u_n \sim u^{fl}_n$

The first fixed point is ultraviolet (UV) stable for $ 0 < \epsilon < 2 $
and infrared  (IR) stable for any other value of $\epsilon$.
For the second fixed point the stability properties are, of course,
opposite. For UV (IR) stable we mean that the fixed point is asymptotically
approached by starting from any initial condition and  by
iterating the ratio-map (2.2) forward (backward). From a physical point of
view,
a forward (backward) iteration of the map (2.2) means a static
cascade of fluctuations
from small (large) scales to large (small) scales.
 In the GOY model, the UV stability is the relevant one, since
 one has a direct cascade of energy or of enstropy.
  As far as the main dynamical mechanism driving
the time evolution of eqs. (1.2) is a forward cascade of energy (like in
$3d$ turbulence), that is for $0<\epsilon <2$
  we expect that the system spends a relevant  fraction of total time nearby
the $K41$-like static solutions. The aim of this paper consists in
quantifying this statement.

Let us stress, also, the importance of the parameter $\epsilon$
from a dynamical point of view. To do this, we introduce the
the  total flux of energy, $\Pi_n$, through the $n$-th shell [6]:
$$
\Pi_n = Im \left( k_n u_n u_{n+1} ( u_{n+2} + {(1-\epsilon) \over 2}
u_{n-1})) \right).
\eqno(2.3)
$$
Where in (2.3)
we have written only the terms coming from the nonlinear transfer of
energy. From (2.3) it is reasonable to expect that by increasing the value
of $\epsilon$ from $0$ to $1$ leads to  a  depletion  of the forward
transfer of energy (the coefficient in front to the  smaller-scales coupling
term
goes to zero). Indeed, numerical integration of GOY models with
$0 < \epsilon < 1$ have shown that the main dynamical  effect
is a forward transfer of energy.

On the other hand, by setting, for example,
$\epsilon = 5/4$, the fluxless like point $u_n \sim k_n^{-1}$
 should dominate the dynamics as it is UV stable
while the Kolmogorov-like  is UV unstable.
Numerically, one observes  a reversed  (backward)
transfer of energy.
In fact, the dynamics of GOY models
with $ \epsilon =5/4$  describes the direct enstrophy cascade
 of $2d$ turbulent flows [1,11].
 For $2>\epsilon>1$,
 beside energy, there also exists
another conserved quantity:
 $\Omega_{\alpha}=\sum k_n^\alpha |u_n|^2$ with $\alpha=-\ln_2 (\epsilon-1)$)
  and in the corresponding shell model one observes
  the direct cascade of such generalized enstrophy $\Omega_{\alpha}$.

Expression (2.3) for the flux of energy also clarifies why
the static solution $u^{fl}_n$  is called ``fluxless''.
Whenever two
shells $u_{n+2} $ and $u_{n-1}$ get trapped by this static fixed point
the flux throughout the shell $n$ is completely inhibited, i.e. $\Pi_n =0$
 (a part viscous and forcing terms). As we will see in the following, the
presence of dynamical barriers for the forward cascade of energy is
considered the main cause of the intermittent nature of the dynamical
evolution.
 \bigskip
\medskip
{\bf 3. The transition to chaos in the shell model}
\bigskip

In this section we present a study of the dynamical properties of the GOY model
in the
 ``forward-energy cascade'' range of parameters ($0 < \epsilon < 1$).

Up until now, the model has been studied numerically and analytically
only for $\epsilon=1/2$ [1,2,6,8]. In this case, the most striking result
is that
 the scaling exponents $\zeta_p$, of the structure functions
($<|u_n|^p> \sim { k_n}^{-\zeta_p}$),
 are a non-linear function of $p$, indicating
 the presence of intermittency in the GOY model which
can be described by the multifractal approach [12].
Moreover, the values of $\zeta_p$ (for  $\epsilon=0.5$) are
very similar to that
 measured  in numerical simulations
 and experiments on real fluids.

 It is an open problem  to relate the multifractality in the $3d$
 real space of the energy dissipation to the multifractality
 of the natural probability measure on the attracting set
 for the dynamics in the $2 N$ phase space.

However,  there is no reason to choose the value $\epsilon=1/2$
 for the coefficients
 of eqs (1.2).
A large spectrum of different behaviors
 can arise in the shell model at varying $\epsilon$,
 the control parameter for
  the  backward flow of the energy in the cascade.

It is remarkable that for $0 <\epsilon \le 0.3843..$, there exists
a finite-Reynolds number fixed point (with viscosity and forcing different
from zero)
which is stable and has Kolmogorov-like scaling in the inertial range.

For example, in figs (1a) and (1b) we have plotted the values of the
ratios $q_n$ at the fixed point obtained from a numerical integration
with $\epsilon =0.05$ and $\epsilon=0.37$. Notice, that the numerical solution
coincides exactly with the result predicted by the ``forward'' iteration
of the ratio-map (2.2) in the inertial range (from the forced shell
to the beginning of the viscous range).

 It is interesting to
remark, also, that the scaling at the fixed point is not exactly
Kolmogorov-like ($q_n=1/2 \, \forall \, n$) because of the damped-oscillation
introduced by the fact that the ratio-map (2.2) does not start exactly at its
fixed point. The oscillations are decreasing by increasing $\epsilon$ and
for small $\epsilon$ they  mask completely the presence of the Kolmogorov
scaling unless one considers a much larger number of shells.

This is, obviously, an effect due to the presence
of the infrared boundary conditions (1.3) at small $n$'s. In the last section
we will come back on this issue, by showing how
to define new infrared boundary conditions which minimize this
effect.

By using the numerical algorithm described in appendix 1
it is possible to follow the fixed point (with viscosity and forcing
different from zero)
and to compute its stability matrix even for values of $\epsilon$
where it is unstable.

To take into account the invariance under rotations
 of the fixed point, in the following we analyze
 the modulus $|u_n|$  rather than the complex variable $u_n$.

 In fig 1c,  the Kolmogorov fixed point  $|u_n^{K41}|$
 is shown for $\epsilon=0.3$ (stable) and $\epsilon=0.5$ (unstable).

By looking at the eigenvalues of the
 stability matrix of the fixed point,
we have detected a Hopf bifurcation at  $\epsilon=0.3843$,   since
 a couple of complex conjugate
eigenvalues   have real part which  passes from negative to positive value.
 The fixed point thus becomes unstable and a stable limit cycle appears
 with a period of $T_1 \approx 90$ natural time units (n.u.).

This  limit cycle loose stability at $\epsilon=0.3953$
 and for $0.3953 < \epsilon<0.398$,
 the attracting set is a torus. The two periods
 of rotations $T_1 \approx 90$  n.u. and
$T_2 \approx 8$ n.u..  The motion on the torus can be
 analyzed by the bi-orthogonal decomposition of the signal
 and one observes that the two rotation periods
 are incommensurate with a ratio $T_1/T_2 =12.05...$.

To illustrate the bifurcation mechanism,
 fig 2 shows the eigenvalues of the stability matrix at
 $\epsilon=0.396$ (immediately after the first transition)
 where there is one couple of
 complex conjugate eigenvalues with positive real part  and at
 $\epsilon=0.396$, (after the second transition)
where there are  two of such couples.

 At $\epsilon=0.398$ there is a third transition to
 an aperiodic attractor with a positive maximum Lyapunov exponent.
 The transition to chaos thus
 seems well described  by the Ruelle-Takens scenario.

 In fig 3, we show the  bi-orthogonal decomposition [7] of
 a signal of $307.2$ natural time units (n.u.) sampled each $0.6$ n.u.
 which has been obtained from a numerical integration with
 a time-step of $3 \, 10^{-4}$ n.u..
 A Fourier spectrum of  the bi-orthogonal decomposition of the signal
  provides a clear evidence of the passage from one frequency,
 to two frequencies and then to chaos, see fig 3d.

At $\epsilon >0.398$,  the time evolution of the dissipative system (1.2)
 is chaotic and confined on
 a strange attractor  in the $2N$ dimensional phase space.
This fact is a strong evidence that the interaction between shells
 plays a fundamental role in determining the strength of
 the intermittency, and that the correct symmetries
 still leave a large freedom to the system.

Let us now add some comments about these different dynamical regimes, as they
can be understood by the bi-orthogonal analysis.
 We refer to [7] for details on this method from which we
 recall only some notations  for the reader's convenience.
 Let us decompose the modulus of the velocity field as
$$
|u_n(t)|=\sum_{k=1}^N A_k \phi_k(n) \, \psi_k(t)
\eqno(3.1)
$$
where $A_1\ge A_2 \ge \cdots \ge A_n >0$ and the $\phi_k$, $\psi_k$ are
 orthonormal functions. The first set of functions, $\phi_k$,
 the so-called {\it Topos}, are the active directions in the configuration
 space while the $\psi_k$,  the so-called {\it Chronos},
 are the corresponding directions in the space of time-series.
 The set of coefficients $A_k$ is the spectrum of the kernel
 operator associated to the signal $|u|$, and is called kinetic
 spectrum, in order to distinguish it from the Fourier spectrum.

First, let us notice that, in all the $\epsilon$-range we have studied, the
 dynamics of $u_n(t)$ in the $2 N$ dimensional phase space
 always evolves in the
neighbourhood of the Kolmogorov fixed point $u^{K41}_n$. this is clearly
 seen from the fact that the first Topos $\phi_1$
 is equal to $u^{K41}$, for any value of $\epsilon$ and
 the orbits stay in a narrow band in the normal direction
 to $\phi_1$ (since $A_n << A_1$ for $n \neq 1$).
All the shell structures that are present are surprisingly stable
 at varying $\epsilon$. As the first nine Topos
 are almost independent of $\epsilon$, only some re-ordering
 occurring in our $\epsilon$ range, we may thus conclude
 that the system essentially lives in a elongated ellipsoid
 inside a space of reduced dimension for most of the time.
 In  [7] one can find an explicit estimation of the time spent
 in that part of the phase space. Nevertheless, the short time spent in the
remaining directions
 of the configuration space is important for
 the mechanism of energy transfer from large scales to the viscous
 small scales. This is shown by the fact that only the last Topos,
 when the latter are ordered by decreasing energy, have support in the
direction of the last shells $19 \, > \, n \ge 16$.

Concerning the inertial range, the most important feature of the dynamics is
that it always takes place along the fixed global structures
of these shells (``coherent structures'').
 It never separates the larger Fourier modes
 from the small ones, that is to say that, during the energy transfer, the
inertial shells are simultaneously and coherently excited.
 It is also easily seen from the shape of the Topos, that the periodicity
three in $n$ plays an important role in the organization of the
energy transfer, thus supporting the analysis made in sect. 2.
 More specifically, for $0.384 \le \epsilon \le 0.394$, all the dynamics,
 included the velocity on the forced shell $u_4$,
 is locked by the fundamental frequency of the circle.
 For larger $\epsilon$-values, up to  $\epsilon=0.395$, the
 shape of the circle is so deformed that a set of new frequencies appears,
 for which the linear approximation around the fixed point is no longer valid.
 In this case, the transfer of energy takes the form of a saw-teeth, a
phenomenon which is reminiscent of heat transfer observed in experiments of
plasma physics [7]. However, due to the lack of smoothness of the orbit, it is
possible that the long range simulation is affected by numerical instabilities
appearing for these particular values of $\epsilon$ close but smaller than the
second bifurcation point $\epsilon=0.3953$.
 For $\epsilon>0.3953$, thanks to the bifurcation to a torus, the energy
transfer is re-organized by the birth of a new frequency, which is able to lock
on the harmonics of the old circle. Indeed, the projection of the dynamics onto
the planes spanned by one of the topos $\phi$
 supported in the infrared region (the
first shells) and each of the $\phi$ supported in the inertial range,  are
circles
 of quasi-periodic motion, as shown in fig 3. Notice that the slopes of $\ln
A_k$, as function of $k$, computed in the ``kinetic inertial range''
 (the part of the kinetic spectrum which is linear in a log-linear plot)
 varies according the bifurcations. It grows after each bifurcation, leading to
a concentration of energy in the first structures of the kinetic inertial range
and then falls at the new bifurcation, showing a new arrangement for the
distribution of the energy inside the kinetic spectrum.

After each of the bifurcations, we also observe a re-ordering of  the
structures
 $\phi_k$, $\psi_k$, and the energy of the structures with
support in the inertial range of the Fourier spectrum always increases.
Finally,
after the torus become unstable, the slope continuously decreases.
 This tendency is compatible  with the route to chaos
 followed by the system at increasing $\epsilon$  (also observed
 in turbulent flows [7]).
As shown in [7], bifurcations take place when certain crossing of the
eigenvalues are present (degeneracy), giving rise to rotations of the space
and time eigendirections in the degenerate eigenspaces.

This is the reason why the entropy of the bi-orthogonal decomposition,
 defined
as
$$
H(|u_n|)=- {1 \over \ln N} \sum_k p_k \ln p_k
\eqno(3.2)
$$
where
$$
p_k={|A_k|^2 \over \sum_k |A_k|^2}
\eqno(3.3)
$$
is a powerful tool for detecting the bifurcations, as one can see in fig 4.
 In order to get a good bifurcation diagram, as it essentially
concerns the kinetic inertial range, we restrict the sum in (3.2) to the
 shell range $n_i \le k \le n_2$. Depending of $\epsilon$,
 $n_1=2$ or $3$ whereas $n_2$ varies from $10$ to $16$
 in the GOY model with $N=19$ shells.  The difference $n_2-n_1$
 grows at increasing $\epsilon$.

The variation of the entropy as function of $\epsilon$ is well understood from
the simultaneous occurrence of degeneracy (the increasing the degree of
equidistribution of the weights $p_k$) and from the exponential decay of the
$A_k$ in the inertial kinetic range.  These two phenomena explain
 the tendency of  the entropy to grow and the occurrence of its local maxima or
minima.

Our physical interpretation is that
when the probability of having a backward energy transfer
 is not large enough,
 the system is able to transfer energy in the most efficient way
 via a non-intermittent cascade.  Above the threshold
 $\epsilon=0.398$ for the transition to chaos,
 backward transfer are so efficient that they are
 able to stop this type of transfer. As a consequence
 the system may charge energy on the first shells.
 During a charge,  one observes a time varying scaling,
 i.e. the velocity $|u_n| \sim k_n^{-h(t)}$
 has an `instantaneous'  scaling exponent $h(t)$ which
  increases  from $1/3$ toward
 larger, and more laminar, values.  At a certain instant, the variables
 $|u_n|$ (with $n$ in the inertial range) become so small
 that viscosity is
 comparable to non-linear transfer
 and dissipate energy directly in the inertial range.
 Then there is a sudden burst which corresponds to a discharge
 of the energy accumulated in the first modes. This is
 a completely different way of dissipating energy,
 which could give origin to multifractality.

The charge-discharge scenario for intermittency has a
 counterpart in the Lyapunov analysis of the shell model, where only
 few degrees of freedom seem to be relevant for the chaotic properties
 of the system.
 Although the Lyapunov dimension of the attractor
 (at least at $\epsilon=0.5$) is proportional to
the total number of shells of the GOY model [1],
 only few Lyapunov exponents
 are positive and there is a large fraction of almost zero
  Lyapunov exponents.
 By an analysis of the Lyapunov eigenvectors, it can be shown that
 they correspond to marginal degrees of freedom which
 concentrate on the inertial range of wavenumbers.
 There are only few degrees of freedom which are chaotic in a very
intermittent way. In fact,
 during the charge, the  energy dissipation stays very low,
 and the instantaneous maximum Lyapunov exponent is almost zero.
 When there is an energy burst,
there is also a  large chaoticity burst,
 i.e. a very large value of the instantaneous maximum Lyapunov exponent,
 with a localization of the corresponding eigenvector on the dissipative
 wavenumbers at the end of the inertial range [2].
 These results have important physical implications
 on the predictability problem which have been discussed in ref [9].

The existence of few active degrees of freedom, in a sea
 of marginal ones, suggests that,
  at least for the dynamics of some global observables,  an appropriate
 one-dimensional map  could capture the essence of the dynamics.
 To verify this idea,  we choose a variable
 which can be interpreted as the  local  singularity,
 or instantaneous scaling exponent  of velocity, that is
$$
 h(t)= {1\over 3} \, {1\over (N-13)} \sum_{7}^{N-7} \ln_2 |u_n/u_{n+3}|.
\eqno(3.4)
$$
The Kolmogorov scaling corresponds to $h=1/3$, and
a laminar signal has $h=1$.
 The choice of the  ratio $q_n=u_{n+3}/u_{n}$
is intended to minimize the effect of period three oscillation proper of
the fixed point structures, taking into account the results
 of section 2.

 We numerically find that for $\epsilon<0.385$
 the local singularity has the constant value
 $h=1/3$ up to an error smaller than $10^{-2}$, as expected.

In figure 5, one sees that at $\epsilon=0.396$
 (the dynamics evolves on a torus)
 the scaling exponent $h(t)$ has very small oscillations
 with two characteristic frequencies around $h=1/3$.

  At increasing $\epsilon$,
the signal $h(t)$ become less and less regular,
 with a  broadening of the probability distribution
 of $h$, as  shown respectively in figs 6a and 6b for $\epsilon=0.42$ and
 figs 7a and 7b for $\epsilon=0.5$

 The maximum scaling exponent $h_{max}\approx 1$
 in both cases, while the minimum one, $h_{min}$,  decreases
 with $\epsilon$.
Note that a value  $h(t)<1/3$ corresponds to a velocity
 field more singular than the one given by the Kolmogorov
 scaling. Such a instantaneous scaling exponent
 is realized during the fast  energy burst
 due to the discharge, while during the charge
 the $h$-value slowly fluctuates in an almost regular way
 around $h\approx 1/3$ and eventually increases
 from $h \approx 1/3$ up to $h \approx 1$.
 We can thus hope to describe
 the most relevant features of the
 dynamics by  looking at the one-dimensional map
 $h(t+\delta t)$ versus $h(t)$  with an appropriate time delay $\delta t$,
  which is shown in figs 6c and 7c
  for $\epsilon=0.42$ and   $\epsilon=0.5$
 It has the typical form of  a map of the Pomeau-Manneville type.
 The channel close to the diagonal is due
 to the charge periods while the relaminarization
 corresponds to a fast  energy burst (the discharge
 process) when a small $h(t+\delta t)$
 follows a rather large $h(t)$.

A further complication arises since we are dealing with a dynamical system
 with many  degrees of freedom. Roughly speaking, the majority of them
 acts as a noisy term  which induces vertical
 (temporal) oscillation on on the one-dimensional map.
 A picture close to the real mechanisms
 that are present in the model, seems therefore to be a
 ``1.5''-dimensional map.
 This will permit to include, more accurately, the shell-time
structure of the symmetries
 that govern the dynamics of the energy transfer. However it is reasonable
 to expect that their statistical effect {\it on the mean quantities}
 is not very important,
 at least  near the transition to chaos.
 Therefore, we have studied the two  cases $\epsilon=0.42$
 (slightly above the transition) and $\epsilon=0.5$ (the usual value
 for the shell model).
  One sees that
the laminar channel of the 1d map
 becomes fatter at increasing $\epsilon$, but the
 relaminarization mechanism is robust.
  As it is well known, the dynamical behavior of $h(t)$ may be
 very well affected by ``random'' oscillations
 of the one-dimensional
 map $y=h(t+\delta t)$ versus $x=h(t)$, close to the diagonal
 $x=y$.
  In particular, this mechanism may also be responsible for the broadening
of the probability distribution of the instantaneous scaling exponent $h$
 at increasing $\epsilon$.
  In practice, the presence of many marginal
 degrees of freedom is revealed by ``random'' oscillations in the
 form of the one dimensional map,  without consequences for
 the qualitative picture.

 It is an open issue to decide whether such a dynamical mechanism
 is relevant to describe the intermittency of real turbulent flow.
 \bigskip
\medskip
\noindent {\bf 4.  A modified GOY model}
\bigskip

To quantify the effect of the intermittent ``charge-discharge'' mechanism
on the scaling exponents $\zeta_p$ it is essential to have an inertial range
as huge as possible and to minimize
non-universal effects due to the infrared  and ultraviolet
 boundary conditions.
In order to have ``ideal'' IR boundary conditions we have to slightly
modify the equations of motion for the first two shells. In this way,
it is possible to  {\it oblige}
the system to  develop a scaling behavior also in the infrared region
 (the first shells) in order to avoid
small deviations of the structure functions
 with respect to  the Kolmogorov prediction $\zeta_p = p/3 $,
 which  could arise as an artefact of the forcing imposed on
 the fourth shells and
of the infrared boundary conditions (1.3).

To show it, let us define a new GOY model which is exactly
equal to the old one but for the following two facts:

\item{(1)} the forcing is moved to the first shell,
\item{(2)} the parameters of the two  equations for $u_1$ and $u_2$ are changed
in the followings:
$$
a_1 \rightarrow 2-\epsilon \,; \,\,\,\,\,
 {\rm instead \,\, of }\,\, a_1=1 \,\,
{\rm   (old \,\, GOY)},
$$

$$
b_2 \rightarrow  -  1  \,; \,\,\,\,\, {\rm \,\, instead\,\,
 of }\,\, b_2= -\epsilon \,\,\,\,
{\rm  (old\,\, GOY)}.
$$
By this choice the requirement of energy conservation (1.4)
 is satisfied, since
 $a_1+b_2+c_3=0$, and the first two equations of (1.2) become:
$$
({d\over dt}+\nu k_1^2 ) \ u_1 \ =
 i \, (2-\epsilon) k_1 \, u^*_{2} u^*_{3} + \ f,
\eqno (4.1)
$$
$$
({d\over dt}+\nu k_2^2 ) \ u_2 \ =
 i \, k_2 \, (  u^*_{3} u^*_{4} - {1 \over 2} \, u^*_{3} u^*_{1}).
\eqno (4.2)
$$

The rationale for the first request is obvious, while the second change
allows us to have an inviscid static fixed point which is at the
fixed point of the map (2.2) for any $n$'s, and therefore the scaling
of the static solutions
 is exactly $u_{n+3}/u_{n} \equiv 1/2, \, \, \forall \,n$.

For example, an inviscid  static solutions will have:
\item{$\bullet$} $(2-\epsilon) k_1 \, u^*_{2} u^*_{3} = -i \,f\, \, $ from the
first equation.
\item{$\bullet$} $u_4=1/2 u_1 \rightarrow q_1 =1/2\, \,$
from the second equation.

Therefore $q_n =1/2 \, \, \forall \,n$,
because the first iteration is already at the UV
stable fixed point of the map (2.2).

For such a class of modified GOY model the static solutions have
exactly $\zeta_p = p/3 $. The dynamical properties are not modified
(energy is still conserved if $\nu=f=0$) and the transition to chaos
follows the same route described above. The advantage is that
now we have a  scaling behaviour which is not affected
from non-universal infrared boundary effect.

If the intermittent mechanism described in the previous section
affects the scaling laws we expect that the beginning of the infrared range
should be much more sensible to the presence of charging process than
the final zone of the inertial range. Indeed,
the probability for a shell nearby the viscous range to be uphill with
respect to a
 barrier of energy is evidently minor than that one of a shell nearby
the forcing zone.  Therefore, small scales are most of the time laminar
or Kolmogorov-like, while large scales are most
of the time in a charging highly-unstable status.

Looking at the scaling laws immediately after the chaotic transition
($\epsilon =0.42$), we have found an interesting
trend of the structure functions to be dominated by the Kolmogorov
scaling by going toward small scales.

To
detect a possible changing
of slope along the inertial range we have used
``local scaling exponents'': $\zeta_p(n)$ [13]. Local scaling exponents are
defined by choosing a fixed length, say $9$ shells, over which fitting
the scaling behavior  of structure functions and then by moving
the analyzed
range of shells
from the infrared region to the dissipation range. With this
definition $\zeta_p(n)$
means the results of the fit performed on the structure functions of order $p$
in the range of $9$ shells centered at shell $m$: $  n-4<m<n+4$.

 In figs 8a and 8b we have plotted
the results for $\zeta_1(n)$ and $\zeta_8(n)$.
 We have used a modified GOY model
with $27$ shells in order to  increase the total length of the inertial range.

 From  fig.s 8 is
possible to see that these ``local scaling exponents''  become
more and more Kolmogorov-like by going toward
the viscous range. In order to improve the quality of our fit
we have used a technique introduced by Benzi et al. [14] called
 Extended-Self-Similarity (ESS). ESS has proved to be efficient
in minimizing finite-size effect and non-universal character
in structure functions. The main idea consists in choosing one
structure functions as reference and then studying the scaling
properties of all other structure functions versus that reference-one.

This trend toward K41 scaling seems to us in agreement with the
previous intermittent picture, small scales are dominated
by laminar or Kolmogorov scaling, while large scales
are most of the time
 more turbulent then a K41 solutions due to the charging process.

 From preliminary data, the same effect seems to be absent for larger
 values of $\epsilon$ (such as the standard value $\epsilon=0.5$)
 which lead to more chaotic systems.
It is an open question whether   the same trend would be present,
by taking a number of shells large enough.
 \bigskip
\medskip
{\bf 5. Conclusions}
\medskip
We have studied the transition to chaos in the GOY shell model,
 at varying the parameter $\epsilon$ related to the  strength of
 backward energy flow.
We thus observe the passage from a stable fixed point
(corresponding to the Kolmogorov non-intermittent energy cascade) toward
 a chaotic attractor (corresponding to the  intermittent cascade)
through the Ruelle-Takens scenario.
We provide a numerical evidence that the
 strange attractor which has a large fractal dimension
 remains close to the  (now unstable) manifold possessing Kolmogorov  scaling.

Immediately above the threshold for chaos, we are able to show that
the physical mechanism of the intermittency of energy dissipation,
 is due to  a charge-discharge mechanism which
 can be described by  a  one-dimensional map.
This is a consequence of the presence of  few `active'
 degrees of freedom
 while the remaining marginal degrees of freedom
 (responsible for the high dimensionality of the attractor)
 have a sort of noisy effect on the one-dimensional map.
The map is of the Pomeau-Manneville type where
 the channel close to the diagonal is related
 to the charge periods while the relaminarization
 corresponds to a fast  energy burst (the discharge
 process).

We have also introduced a modified shell model where
there is  a good scaling behavior even
 in the infrared (small wave-number) range. The presence of a huge range
of  scaling shells
allows us to study in detail the possible presence of deviations
to the usual power law scaling.
 We find that for the GOY model in the ``weak'' chaotic region
($\epsilon=0.42$) the structure functions tend to become Kolmogorov-like
by decreasing the analyzed scales. This  could be an indication
 that multifractal corrections disappear in the limit of large Reynolds number,
 at least for $\epsilon$ slightly above the transition to chaos.
It is very difficult to decide by numerical experiments
 if such an effect is present at the usual value $\epsilon=0.5$,
because one should consider very high Reynolds that is
 a very large number of shells.
 It still remains  an open problem
 to understand whether the charge-discharge intermittency
 described in this paper  might be compatible with
 the Kolmogorov scaling laws, or it  brakes a global scaling
 invariance leading to multifractality, as commonly believed on
 the basis of numerical experiments [2-6] and analytic calculations [8].
 \bigskip
\bigskip
{\bf Acknowledgments}
We thank Daniele Carati for many interesting discussions.
L. B. was partially supported by a Henri Poincar\'e
fellowship
 (Centre National de la Recherche Scientifique and Conseil
G\'en\'eral des Alpes Maritimes) and by the ``Fondazione Angelo della Riccia''.
G.P. is grateful to the C.P.T. de Luminy for warm hospitality.
 \vfill\eject
\centerline{\bf Appendix}
\bigskip
In this appendix we show the numerical algorithm for the search of the
 Kolmogorov fixed point at $\epsilon >  0.3843$ where it is unstable.
Let us denote by
$$
{ dU  \over dt}=F(\epsilon,U), \eqno (A1)
 $$
 the system (1.2) considered as a real $2 \, N$ dimensional
 system where  $U_n=Re(u_n)$ and $U_{n+N}=Im(u_n)$ with $n=1,\cdots,N$.
   The point $U^{K41}(\epsilon)$ is a fixed point of (A1) if
$F(\epsilon, U^{K41})=0$.

In order to determine the value of the fixed point for $\epsilon +
 \delta \epsilon$, we make the observation, which stems
 from numerical simulations
 in a range of $\epsilon$-values where the fixed point is stable,
 that $ U^{K41}(\epsilon)$ moves very slowly with $\epsilon$.
 We can thus expand up to the first order in $\delta \epsilon$
 the relation
$$
F(\epsilon+\delta \epsilon, U^{K41}(\epsilon+\delta \epsilon))=0.
 \eqno (A2)
$$

As $F$ depends linearly on $\epsilon$ we have the exact relation
$$
 F(\epsilon, U^{K41}(\epsilon+\delta \epsilon))
 +
\delta \epsilon {\partial
 F(\epsilon, U^{K41}(\epsilon+\delta \epsilon)) \over \partial \epsilon}=0.
\eqno (A3)
$$
By assuming that
$$
 U^{K41}(\epsilon+\delta \epsilon)=
 U^{K41}(\epsilon)+V^{(\epsilon)} \delta \epsilon,
$$
and using the fact that
$ F(\epsilon, U^{K41}(\epsilon))=0$, one obtains from (A3)
$$
D F(\epsilon, U^{K41}(\epsilon)) \, V^{\epsilon} \, + \,
 {\partial
 F(\epsilon, U^{K41}(\epsilon))\over \partial \epsilon}=0,
\eqno (A4)
$$
where $DF (\epsilon, U^{K41}(\epsilon))$ is the the stability matrix
 of the system (1.2) calculated at $U^{K41}(\epsilon)$.
 this equation can be solved in $V$ and reads
$$
U^{K41}(\epsilon+\delta \epsilon)=
 U^{K41}(\epsilon)-\delta \epsilon \,
[D F(\epsilon, U^{K41}(\epsilon))]^{-1} \,
 {\partial
 F(\epsilon, U^{K41}(\epsilon))\over \partial \epsilon}.
\eqno (A5)
$$
 The two matrices $DF$ and $\partial F/\partial \epsilon$
 are obtained by a direct numerical calculation.
 In this paper we have iterated $A5$ with $\delta \epsilon=10^{-4}$,
 starting from a stable fixed point $U^{K41}{\epsilon_0=0.2}$
 which has been obtained by a long numerical integration
 of the shell model. The stability matrix $DF$ is then
 found and diagonalized at the $\epsilon$'s  of interest
(see fig.s 2 for $\epsilon=0.386$ and $\epsilon=0.396$).
 \vfill\eject
\noindent
{\bf References.}
\bigskip
\baselineskip=8pt
\parskip=5pt
\item{[1]}
M. Yamada and K. Okhitani, J. Phys. Soc. of Japan {\bf 56}, 4210 (1987);
 Progr. Theo. Phys. {\bf 79}, 1265 (1988);
 Phys. Rev. Lett. {\bf 60}, 983 (1988).
\item{[2]}
M.H. Jensen, G. Paladin and A. Vulpiani, Phys.Rev.A
{\bf 43}, 798 (1991).
\item{[3]}
A.M. Obukhov, Atmos. Oceanic. Phys. {\bf 7}, 41 (1971);
A.M. Obukhov, Atmos. Oceanic. Phys. {\bf 10}, 127 (1974);
V.N. Desnyaski and E.A. Novikov, Prikl. Mat. Mekh. {\bf 38} 507 (1974)
\item{[4]}
E.B. Gledzer, Sov. Phys. Dokl. {\bf 18}, 216 (1973).
\item{[5]}
E.D. Siggia, Phys. Rev.A {\bf 15}, 1730 (1977);
E.D. Siggia, Phys. Rev.A {\bf 17}, 1166 (1978);
R.M. Kerr and E.D. Siggia, J. Stat. Phys. {\bf 19}, 543 (1978);
R. Grappin, J. Leorat and A.  Pouquet, J. de Physique {\bf 47}, 1127
(1986).
T. Bell and M. Nelkin, Phys. of Fluids {\bf 20} 345 (1977).
\item{[6]}D. Pisarenko, L. Biferale, D. Courvasier, U. Frisch and M. Vergassola
 Phys. of Fluids {\bf A5}, 2533 (1993).
\item{[7]}N. Aubry, R. Guyonnet and R. Lima, J. Stat. Phys.
 {\bf 64} n. 3/4 (1991); ibidem {\bf 67} n. 1/2 (1992);
 Th. Dudok de Wit, R. Lima, A.-L. Pecquet, J.-C. Vallet
 to appear in Phys. of Fluids {\bf B} (1994);
N. Aubry, F. Carbone, R. Lima and S. Slimani to appear in J. Stat. Phys.
 (1994)
\item{[8]} R. Benzi,  L. Biferale and G. Parisi,  Physica {\bf D65}, 163 (993)
\item{[9]}
A. Crisanti, M.H. Jensen, G. Paladin and A. Vulpiani, Phys. Rev. Lett.
{\bf 70}, 166 (1993).
\item{[10]} L. Biferale, M. Blank and U. Frisch, to appear in
J. Stat. Phys. (1994).
\item{[11]} P. Frick and  E. Aurell,  Europhys. Lett. {\bf 24} 725 (1993).
\item{[12]} G. Paladin and A. Vulpiani Phys. Rep. {\bf 156}, 147 (1987).
\item{[13]} S. Grossmann and D. Lohse, ``Universality in fully
developed turbulence'', Preprint of the University of Chicago, (1993).
\item{[14]} R. Benzi, S. Ciliberto, C. Baudet, R. Tripiccione,
F. Massaioli and S. Succi, Phys. Rev. E, {\bf 48} R29 (1993).
\vfill\eject
\noindent
{\bf FIGURE CAPTIONS}
 \bigskip
\baselineskip=8pt
\parskip=5pt
\item{Fig 1a} Values of the ratios $q_n$ at the fixed point
of equations (1.2) with $\epsilon=0.05$ and with superimposed the values
predicted by the ratio-map (2.2). Circles are the outputs from the numerical
integration, where squares correspond to the ratio-map values. The straight
line correspond to the exact K41 scaling ($q_n=1/2 \,\forall \, n$)
\item{Fig 1b} The same as in figure (1a) but with $\epsilon=0.37$.
\item{Fig 1c}
 Kolmogorov fixed point $\ln_2 |{u^{K41}}_n|$ versus $n$
 for $\epsilon=0.3$ (dashed line) and $\epsilon=0.5$ (solid line).
\item{Fig 1d} Topos $\phi_1$
 versus $n$ for $\epsilon=0.3$ (dashed line) and $\epsilon=0.5$ (solid line),
 given by the bi-orthogonal decomposition of
a signal obtained from a numerical integration
 of $307.2$ n.u.  after a transient of $3900$ n.u.
starting from
 an initial condition close to the Kolmogorov  fixed point.
These initial conditions are also used to obtain figs 4, 5, 6 and 7.
 \item{Fig 2}
 Imaginary versus real part of the
eigenvalues of the stability matrix of the Kolmogorov fixed point at
 $\epsilon=0.386$ (limit cycle) and at  $\epsilon=0.396$ (torus).
  \item{Fig 3}
Bi-orthogonal decomposition of
 a signal obtained from a numerical integration
 of $307.2$ n.u.  after a transient of $3900$ n.u.. In figs
 3a, 3b and 3c the time unit of the horizontal axis is 0.6 n.u..
 \item{Fig 3a}
 Three-dimensional plot of the torus obtained by the Chronos $\psi_8$,
 $\psi_9$ and $\psi_3$.
 \item{Fig 3b}Oscillations of Chronos $\psi_3(t)$
  with period $T_1 \approx 90$ n.u..
 \item{Fig 3c}Oscillations of Chronos $\psi_9(t)$
  with period $T_2 \approx 8$ n.u.,
 modulated by the first harmonics of period $T_1$
  \item{Fig 3d} Semi-log plot  of the
 Fourier power spectrum of Chronos $\psi_9$.
 \item{Fig 4}
Entropy of the bi-orthogonal decomposition
 versus $\epsilon$  where $n_1=3$ and $n_2=N=19$.
 For each point, the entropy is obtained from
 a numerical integration
 of $307.2$ n.u.  after a transient of $6000$ n.u.
starting at the corresponding
Kolmogorov  fixed point $u^{K41}(\epsilon)$.
These initial conditions are also used to obtain figs 1d, 5, 6 and 7.
 \item{Fig 5}
 Instantaneous scaling exponent $s(t)=3 \, h(t)$ as function of time
 at  $\epsilon=0.396$ (torus). The Kolmogorov scaling corresponds to $s=1$.
 \item{Fig 6a}  Instantaneous scaling exponent $s(t)=3 \, h(t)$
 as function of time,  at  $\epsilon=0.42$. Note that
 a laminar velocity field has $s=3$, and the Kolmogorov fixed point $s=1$.
  \item{Fig 6b} Probability distribution of  the
  instantaneous scaling exponent $s=3 \, h$
 at  $\epsilon=0.42$.
 \item{Fig 6c}
 One-dimensional map obtained by plotting
 $s(t+\delta t)$ versus $s(t)$  with $s(t)=3 \, h(t)$, $\delta t=0.6$ n.u.
  for $\epsilon=0.42$.
 \item{Fig 7a}  Instantaneous singularity $s(t)=3 \, h(t)$
  as function of time
 at  $\epsilon=0.5$.
 \item{Fig 7b} Probability distribution of  the
  instantaneous singularity $s=3 \, h$
 at  $\epsilon=0.5$.
 \item{Fig 7c}
 One-dimensional map obtained by plotting
 $s(t+\delta t)$ versus $s(t)=3 \, h(t)$  with $\delta t=0.6$ n.u.
  for $\epsilon=0.5$.
 \item{Fig 8a} Local scaling exponent for structure function of
order $1$. Notice the trend toward the K41 value: $\zeta_1=1/3$
 by decreasing the set of analyzed scales in the inertial range.
 \item{Fig 8b} The same as in figure (8a) but for the structure
functions of order $8$ (here the k41 value  corresponds
to $\zeta_8= 8/3 = 2.666..$).
\end